\documentclass[
superscriptaddress,
twocolumn,
showpacs,preprintnumbers,
amsmath,amssymb,
aps,
prl,
]{revtex4-1}

\usepackage{graphicx}
\usepackage{dcolumn}
\usepackage{bm}

\begin{document}


\title{VI$_{3}$: a 2D Ising ferromagnet}


\author{Ke Yang}
 \affiliation{Laboratory for Computational Physical Sciences (MOE),
 State Key Laboratory of Surface Physics, and Department of Physics,
 Fudan University, Shanghai 200433, China}

\author{Fengren Fan}
 \affiliation{Laboratory for Computational Physical Sciences (MOE),
 State Key Laboratory of Surface Physics, and Department of Physics,
  Fudan University, Shanghai 200433, China}

\author{Hongbo Wang}
 \affiliation{Laboratory of Advanced Materials Physics and Nanodevices, School of Physics and Technology, University of Jinan, Jinan 250022, China}

\author{D. I. Khomskii}
\affiliation{Institute of Physics II, University of Cologne, 50937 Cologne, Germany}

\author{Hua Wu}
\email{Corresponding author. wuh@fudan.edu.cn}
\affiliation{Laboratory for Computational Physical Sciences (MOE),
 State Key Laboratory of Surface Physics, and Department of Physics,
 Fudan University, Shanghai 200433, China}
\affiliation{Collaborative Innovation Center of Advanced Microstructures,
 Nanjing 210093, China}

%
%


\begin{abstract}

Two-dimensional (2D) magnetic materials are of great current  interest for their promising applications in spintronics. Here we propose the van der Waals (vdW) material VI$_3$ to be a 2D Ising ferromagnet (FM), using density functional calculations, crystal field level diagrams, superexchange model analyses, and Monte Carlo simulations.
The $a_{1g}$$^1$$e'_{-}$$^1$ ground state in the trigonal crystal field gives rise to the 2D Ising FM due to a significant single ion anisotropy (SIA) and enhanced FM superexchange both associated with the $S_z$=1 and $L_z$=--1 state of V$^{3+}$ ions.
We find that a tensile strain on the VI$_3$ monolayer further stabilizes the $a_{1g}$$^1$$e'_{-}$$^1$ ground state, and its Curie temperature ($T_{\rm C}$) would increase from 70 K to 90-110 K under a 2.5-5\% tensile strain. Moreover, we suggest a group of spin-orbital states with a strong SIA which may help to search more 2D Ising magnets.

\end{abstract}

\maketitle



Two-dimensional (2D) crystals with intrinsic magnetism have been of great interest since the experimental achievements of the atomically thin CrI$_{3}$~\cite{Huang_2017} and Cr$_{2}$Ge$_{2}$Te$_{6}$~\cite{Gong_2017} flakes by exfoliation of bulk crystals. Through magneto-optical Kerr effects, the CrI$_{3}$ monolayer and Cr$_{2}$Ge$_{2}$Te$_{6}$ atomic layers have been demonstrated to be a ferromagnet (FM) with out-of-plane spin orientation. More recently, exfoliated Fe$_{3}$GeTe$_{2}$ monolayers have $T_{\rm C}$ above room temperature via ionic gating~\cite{Deng_2018}. Those 2D FMs provide unique opportunities for understanding, exploring and utilizing novel low-dimensional magnetism. Due to the thickness of  one or few monolayers, one may be able to control the 2D magnetic properties by applying weak magnetic field~\cite {Huang_2018_nature,Jiang_2018,Sun_2019}, electric field~\cite {Huang_2018_nature,Jiang_2018}, doping~\cite{Deng_2018,Huang_2018_jacs}, or heterostructure~\cite{Gibertini_2019,Song_2018}. This flexibility causes enormous excitement about their promising applications in spintronics.

Very recently, VI$_{3}$ emerges as a new 2D material with a similar van der Waals (vdW) layered structure as CrI$_{3}$ which is currently under extensive study~\cite{Huang_2017,Huang_2018_nature,Jiang_2018,Sun_2019,Huang_2018_jacs,Gibertini_2019,Song_2018,Mcguire_2015,Wang_2016,Lado_2017,Kim_2019}. The bulk material of VI$_{3}$ has been studied by several groups~\cite{Kong_2019,Son_2019,Tian_2019,Yan_2019,Gati_2019,Dole_2019,Liu_2020}, and it is found to be an interesting FM insulator with $T_{\rm C}\approx$ 50 K and the easy magnetization $c$-axis~\cite{Kong_2019,Son_2019}. Note that CrI$_3$ has a closed $t_{2g}$$^3$ shell for the octahedral Cr$^{3+}$ $S$=3/2 ion. Therefore, its orbital singlet produces no single ion anisotropy (SIA), and its finite perpendicular magnetic anisotropy comes from the exchange anisotropy caused by the spin-orbit coupling (SOC) of the heavy I $5p$ orbitals and their hybridization with the Cr $3d$~\cite{Lado_2017,Kim_2019}. In contrast, VI$_3$ has an open $t_{2g}$$^2$ shell for the $S$=1 V$^{3+}$, and therefore it has room to achieve an orbital moment and a consequent strong SIA due to SOC. Then, VI$_3$ may be an Ising type 2D FM.

In this Rapid Communication, we indeed find that the insulating VI$_3$ vdW monolayer has the $a_{1g}$$^1$$e'_{-}$$^1$ ground state with $S_z$=1 and $L_z$=--1 in the trigonal crystal field. The FM superexchange and the strong perpendicular SIA produce the 2D Ising FM. The $a_{1g}$$^1$$e'_{-}$$^1$ ground state can further be stabilized by a tensile strain, which would enhance $T_{\rm C}$ of the VI$_3$ monolayer from 70 K to 90-110 K under 2.5-5\% strain. Therefore, the VI$_3$ monolayer could be the first real 2D Ising FM insulator, which calls for a prompt experimental verification.


We have carried out density functional theory (DFT) calculations for both bulk and monolayer VI$_3$, using the experimental lattice parameters~\cite{Kong_2019} and the full-potential augmented plane wave plus local orbital code (Wien2k)~\cite{WIEN2K}, see more computational details in Supplemental Material (SM).
To account for the electron correlation of the narrow V $3d$ bands, the local-spin-density approximation plus Hubbard U (LSDA+U) calculations were performed using Hubbard U=4.0 eV and Hund exchange J$_{\rm H}$=0.9 eV~\cite{Vladimir_1997}. The SOC is included for both V $3d$ and I $5p$ orbitals by the second-variational method. As seen below, the SOC is crucial and any reasonable U (e.g., 2-5 eV) is big enough to open an insulating gap between the SOC split $e'_{\pm}$ states and hence reach the same ground state solution for VI$_3$. Moreover, we have used the crystal field and superexchange pictures to understand the 2D Ising FM in VI$_3$ as detailed below. Furthermore, we have estimated the $T_{\rm C}$ of VI$_{3}$ monolayer using Monte Carlo (MC) simulations on a $6\times6\times1$ spin matrix. At each temperature, 2.4$\times$10$^{7}$ MC steps/site were performed to reach an equilibrium using the Metropolis method~\cite{Nicholas_1949}, and then the specific heat is calculated.

 \begin{figure}[t]
\includegraphics[width=7.5cm]{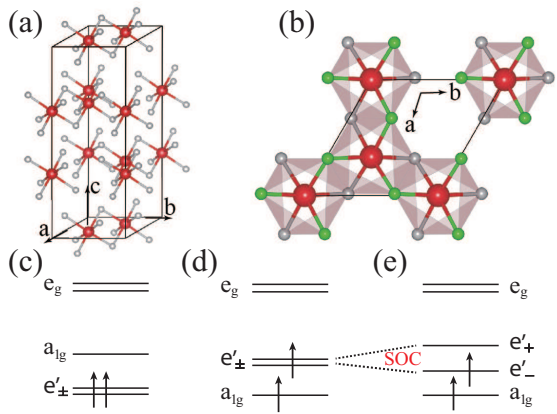}
 \caption{(a) The bulk crystal structure of VI$_{3}$, with V (I) atoms represented by red (grey) balls. (b) The honeycomb V lattice of VI$_{3}$ monolayer, and the green (grey) balls standing for the I atoms above (below) the V layer, forming the edge sharing VI$_6$ octahedra. (c) The $e'_{\pm}$$^{2}$ or (d) $a_{1g}$$^{1}$$e'_{\pm}$$^{1}$ configuration for the V$^{3+}$ $S$=1 ion in the trigonal crystal field level diagrams, with the active SOC effect in (e).
}
\label{fig1:el}
\end{figure}
%


We first consider the situation in the bulk VI$_3$ where some recent experimental results~\cite{Kong_2019,Son_2019} are available for comparison. According to the crystal structure of the bulk VI$_3$ in the $R\overline{3}$ space group~\cite{Kong_2019}, the V ions have a local octahedral coordination but a trigonal crystal field in the global coordinate system, which splits the otherwise degenerate $t_{2g}$ triplet into the $a_{1g}$ singlet and $e_{\pm}'$ doublet, see Fig. 1. Several very recent DFT studies gave conflicting results for the bulk VI$_3$ (either metallic~\cite{Kong_2019,He_2016} or insulating~\cite{Tian_2019,Son_2019}),
but so far no understanding of the perpendicular FM has been available. In the DFT+U framework one may readily get the insulating solution with the $e'_{\pm}$$^2$ double occupation, in contrast to the metallic state $a_{1g}$$^1$$e'_{\pm}$$^1$ with the half filled $e'_{\pm}$ doublet. However, it is the half-filled $e'_{\pm}$ doublet (Figs. 1(d) and 1(e)) which makes the SOC active and may eventually determine the perpendicular FM. For this purpose, we have carried out LSDA+SOC+U calculations for the bulk VI$_3$ to make a direct comparison between  two configurations,  $e'_{\pm}$$^2$ and $a_{1g}$$^1$$e'_{\pm}$$^1$. These two configurations are initialized via the occupation density matrix over the eigen orbitals~\cite{Ou_2015,Ou_2014}, rather than a common representation by the two doublets ($xy$,~$x^2$--$y^2$) and ($xz$,~$yz$) in the trigonal crystal field, which may not find the correct $a_{1g}$$^1$$e'_{-}$$^1$ ground state in the calculations.

Our LSDA+SOC+U calculations show that both insulating solutions can be stabilized, and that the $a_{1g}$$^1$$e'_{-}$$^1$ state, now with $l_z$=--1, is  more stable than the $e'_{\pm}$$^2$ state, by 9.3 meV/fu, as seen in Table I. Both the solutions have the local V$^{3+}$ spin moment more than 1.8 $\mu_{\rm B}$ and the total spin moment of 2 $\mu_{\rm B}$/fu, showing exactly the formal V$^{3+}$ $S$=1 state. Note that the former solution has now a large orbital moment of --1.05 $\mu_{\rm B}$ along the $z$-axis due to filling of the lower $e'_{-}$ ($l_z$=--1) level after the SOC splitting of the $e'_{\pm}$ doublet (Fig.1(e)). In contrast, the fully occupied $e'_{\pm}$$^2$ doublet has no orbital degree of freedom and thus only a very small orbital moment of 0.05 $\mu_{\rm B}$ is induced by the SOC. To prove the essential role of the SOC, we have also calculated the $a_{1g}$$^1$$e'_{+}$$^1$ state with filling of the SOC-split upper $e'_{+}$ ($l_z$=1) level. The resulting total energy rises by 38.7 meV/fu, compared with the $a_{1g}$$^1$$e'_{-}$$^1$ ground state. Then the SOC parameter of the V$^{3+}$ ion $\xi$=38.7 meV is derived, and it is (largely) responsible for the energy lowering of the $a_{1g}$$^1$$e'_{-}$$^1$ ground state, relative to the $e'_{\pm}$$^2$ state where the SOC is almost absent. Note that the SOC parameter $\xi$ is partially enhanced~\cite{Liu_2008,Haverkort_2008} here by the strong SOC of the heavy I atom via the I $5p$-V $3d$ hybridization.

\renewcommand\arraystretch{1.3}
\begin{table}[t]
  \caption{Relative total energies $\Delta$E (meV/fu) by LSDA+SOC+U, local spin and orbital moments ($\mu_{\rm B}$) for the V$^{3+}$ ion. The perpendicular magnetization is assumed in most cases, and the in-plane magnetization is also set for $a_{1g}$$^{1}$$e'_{-}$$^{1}$. The corresponding data for the fully relaxed structures are listed in the round brackets.
}
\label{tb1}
\begin{tabular}{lcccc}
\hline\hline
VI$_{3}$ bulk    &  &    $\Delta$E & M$_{\rm spin}$   & M$_{\rm orb}$  \\ \hline
$a_{1g}$$^{1}$$e'_{-}$$^{1}$ & FM  &  0.0 (0.0)      &  1.89 (1.86)      & --1.05 (--1.00) \\            & AF  &  21.3      &  1.85       & --1.07               \\ \hline
$a_{1g}$$^{1}$$e'_{+}$$^{1}$   & FM  &  38.7      &  1.89       & 1.03        \\  \hline
$e'_{\pm}$$^{2}$     & FM  &  9.3 (11.8)      &  1.83 (1.80)   & 0.05 (0.05) \\
&     AF  &  15.8      &  1.81       & 0.07                \\
\hline\hline
VI$_{3}$ monolayer  &   & $\Delta$E & M$_{\rm spin}$   & M$_{\rm orb}$   \\ \hline
$a_{1g}$$^{1}$$e'_{-}$$^{1}$ & FM      & 0.0 (0.0)    & 1.88 (1.84)  &  --1.08 (--1.04) \\               &AF      & 20.2 (20.0)    & 1.84 (1.80)   &  --1.09 (--1.05)          \\ \hline
$a_{1g}$$^{1}$$e'_{-}$$^{1}$  & FM      & 17.0 (17.6)   & 1.88 (1.84)  &  --0.15 (--0.15)   \\
   (in-plane M)    & AF      & 35.2 (33.9)    & 1.84 (1.80)   &  --0.14 (--0.17)  \\ \hline
$e'_{\pm}$$^{2}$  & FM    & 20.1 (45.5) & 1.82 (1.78)   &  0.05 (0.05)             \\
& AF    & 27.1 (52.5) & 1.80 (1.76)   &  0.07 (0.07)        \\
\hline\hline
 \end{tabular}
\end{table}

Moreover, for both insulating solutions, $a_{1g}$$^1$$e'_{-}$$^1$ and $e'_{\pm}$$^2$, the FM state is more stable than the antiferromagnetic (AF) state, see Table I. While the FM stability against AF is 6.5 meV/fu for the $e'_{\pm}$$^2$ state, it increases a lot to 21.3 meV/fu for the $a_{1g}$$^1$$e'_{-}$$^1$ ground state. This enhanced FM superexchange in the $a_{1g}$$^1$$e'_{-}$$^1$ ground state appears also in the VI$_3$ monolayer and will be explained below. The FM order can further be stabilized by its Ising type magnetism due to the SOC between the $S_z$=1 and $L_z$=--1. The total magnetic moment of about 1 $\mu_{\rm B}$/fu is strongly reduced from the V$^{3+}$ $S$=1 state, and it well accounts for the experimental easy $z$-axis magnetization of 1-1.3 $\mu_{\rm B}$ in the FM insulating VI$_3$~\cite{Kong_2019,Son_2019}.

As VI$_3$ bulk has already the potential to be an Ising FM, its vdW monolayers could well be a 2D Ising FM, with a better tunability. The cleavage energy is calculated to be 0.27 J/m$^{2}$, using DFT+vdW corrections within the Grimme's approach~\cite{Grimme_2006} (see SM for details), and it is well comparable to those for other 2D materials such as CrI$_{3}$ (0.30 J/m$^{2}$)~\cite{Mcguire_2015}, Cr$_{2}$Si$_{2}$Te$_{6}$ (0.35 J/m$^{2}$) and Cr$_{2}$Ge$_{2}$Te$_{6}$ (0.38 J/m$^{2}$)~\cite{Li_2014}. Therefore, an exfoliation of VI$_{3}$ monolayer is likely, and we now switch to the treatment of the VI$_3$ monolayer.

Our LDA and LSDA calculations find a FM metallic solution, and LSDA+U calculations give a FM insulating solution with $e'_{\pm}$$^2$, see SM for details. One may assume that the experimental FM insulating behavior is reproduced by the $e'_{\pm}$$^2$ state. However, a key point is the following: the V$^{3+}$ $e'_{\pm}$$^2$ state has only the pure $S$=1 and a quenched orbital moment, i.e., no SIA which would be very beneficial for the 2D FM. Apparently, such a solution would hardly explain the strong perpendicular FM observed in bulk VI$_3$~\cite{Kong_2019, Son_2019}. Then, one may have to resort, as in the extensively studied CrI$_3$~\cite{Lado_2017,Kim_2019}, to the weak exchange anisotropy due to the SOC of I $5p$ orbital and its hybridization with V $3d$. But actually, VI$_3$ has a more than one order of magnitude stronger SIA, which determines its 2D Ising FM as demonstrated below.

\begin{figure}
\includegraphics[width=5.5cm]{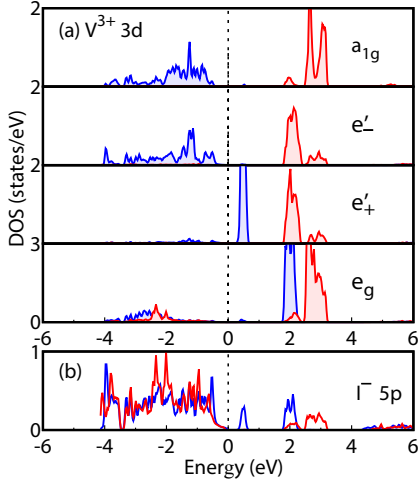}
 \caption{The $a_{1g}$$^{1}$$e'_{-}$$^{1}$ ground state with a strong V $3d$-I $5p$ hybridization by LSDA+SOC+U. The blue (red) curves stand for the up (down) spin. Fermi level is set at zero energy.
 }

\end{figure}

Our LSDA+SOC+U calculations give the insulating $a_{1g}$$^1$$e'_{-}$$^1$ ground state (Fig. 2), with the gap opening due to the electron correlation within the SOC-split $l_z=\pm1$ states, and it is more stable than the $e'_{\pm}$$^2$ solution by 20.1 meV/fu, see Table I. Note that for the partially-filled $t_{2g}$ systems, there is an old and well-known dichotomy (see. e.g. Ref.~\onlinecite{Khomskii_2001}): such ions, on one hand, are Jahn-Teller (JT) active and may distort and fill orbitals according to the JT scenario - here for V$^{3+}$ $3d^{2}$ occupation, giving the $e'_{\pm}$$^{2}$ state. But they can also form the state with unquenched orbital moment and gain extra SOC energy. The above results imply that the SOC effects, from V $3d$ itself and I $5p$ via the $p$-$d$ hybridization (see Fig. 2), indeed favor the $a_{1g}$$^1$$e'_{-}$$^1$ ground state solution.

 \begin{figure}
\includegraphics[width=8.5cm]{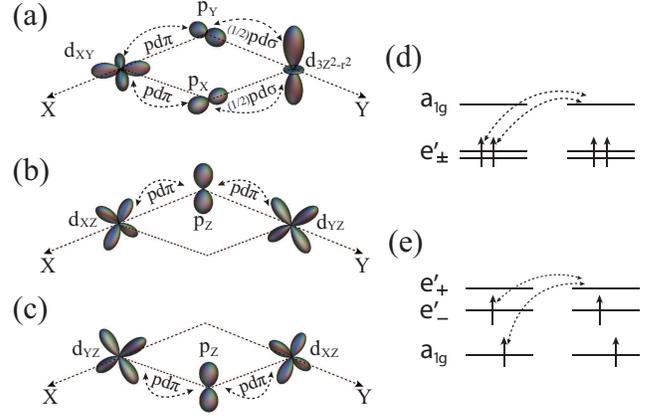}
 \caption{Schematic plot of the hopping channels involved in the (near-) 90$^{\circ}$ FM superexchange: (a) d$_{XY}$ - (p$_{X}$, p$_{Y}$) - d$_{3Z^{2}-r^{2}}$, (b) d$_{XZ}$ - p$_{Z}$ - d$_{YZ}$, and (c) d$_{YZ}$ - p$_{Z}$ - d$_{XZ}$.  Virtual hoppings between the V$^{3+}$ ions with (d) $e'_{\pm}$$^{2}$ and (e) $a_{1g}$$^{1}$$e'_{-}$$^1$.
}
\label{fig1:el}
\end{figure}

Moreover, for the $a_{1g}$$^1$$e'_{-}$$^1$ ground state, the FM state is more stable than the AF state by 20.2 meV/fu, which is very similar to the bulk case, see Table I. In contrast, for the $e'_{\pm}$$^{2}$ state, the FM state is more stable than the AF state only by 7.0 meV/fu. Here we provide a picture to understand the enhanced FM superexchange in the $a_{1g}$$^1$$e'_{-}$$^1$ ground state. Considering the honeycomb lattice of the V$^{3+}$ magnetic ions and the edge-sharing VI$_6$ octahedral network (see Fig. 1), we discuss the near-90$^\circ$ FM superexchange interactions, see Fig. 3.  Here we choose a local octahedral $XYZ$ coordinate system, with the $XYZ$ axes directed from V to neighboring I ions. Then the local octahedral $t_{2g}$ triplet under a trigonal crystal field can be expressed as
\begin{equation}
\begin{aligned}
&a_{1g}=\frac{1}{\sqrt{3}}(XY + XZ + YZ) \\
&e'_+=\frac{1}{\sqrt{3}}(XY + e^{i2\pi/3}XZ + e^{i4\pi/3}YZ) \\
&e'_-=\frac{1}{\sqrt{3}}(XY + e^{i4\pi/3}XZ + e^{i2\pi/3}YZ). \\
\end{aligned}
\end{equation}
The V-V superexchange contains several contributions. In analogy with the well studied CrI$_3$, the main FM contribution comes from the occupied $t_{2g}$ - empty $e_g$ virtual hoppings~\cite{Wang_2016}.  There are also AF processes, due to hoppings between the occupied $t_{2g}$ orbitals. Importantly, in VI$_3$, and in contrast to CrI$_3$, there appear the FM contributions due to hoppings from occupied to empty $t_{2g}$ orbitals; these will turn out to be crucial in enhancing FM contributions for the $a_{1g}$$^1$$e'_{-}$$^1$ ground state.

We schematically illustrate these contributions treating VI$_3$ as a Mott-Hubbard (MH) insulator. Actually for such ligand as I the system may be close to a charge-transfer (CT) regime, see e.g. Ref.~\onlinecite{Khomskii_2001}; this would add extra terms in the superexchange, but this would not change the main conclusions here. In the MH regime the main superexchange occurs due to the effective $d$-$d$ hoppings through the same $5p$ orbital of I ions (in CT case also the hoppings via different orthogonal $5p$ orbitals could enter). The main FM contribution  comes from the effective V-V hopping from each of the above three orbitals to the neighboring empty (real) $e_g$ orbital $3Z^2-r^2$ via the mechanism shown in Fig. 3(a) (the hoppings to $X^2-Y^2$ orbital via two iodine ions cancel due to the signs of the $d$- and $p$-wave functions). According to Fig. 3(a), only the $XY$ component of each of the states in Eq. 1 is active here, and the hoppings from each of them to the $3Z^2-r^2$  orbital of a neighboring V  are equal, and we see that this FM contribution is the same for both the $a_{1g}$$^1$$e'_{-}$$^1$ and $e'_{\pm}$$^2$ configurations.

As mentioned above, also for $t_{2g}$-$t_{2g}$  contributions, the V-V hopping and superexchange via the common single $5p_z$ orbital of two I$^{-}$ ligands, as sketched in Figs. 3(b) and 3(c), are also effective. Then we have the effective hoppings as follows
\begin{equation}
\begin{aligned}
\langle{a_{1g}}|\hat{t}|a_{1g}\rangle=-2t_{0},
~~~\langle{e'_{+}}|\hat{t}|e'_{+}\rangle=\langle{e'_{-}}|\hat{t}|e'_{-}\rangle=-t_{0},\\
\langle{e'_{+}}|\hat{t}|e'_{-}\rangle=-2t_{0},
~~~\langle{a_{1g}}|\hat{t}|e'_{+}\rangle=\langle{a_{1g}}|\hat{t}|e'_{-}\rangle=-t_{0},
\end{aligned}
\end{equation}
where $t_0$ = $t_{pd\pi}^{2}/(3\Delta)$ and $\Delta$ is the charge transfer energy.
Using these hoppings, one can show that the AF contribution due to hopping between occupied $t_{2g}$ orbitals is again the same for both $a_{1g}$$^1$$e'_{-}$$^1$ and $e'_{\pm}$$^2$. This is however not the case for the FM $t_{2g}$-$t_{2g}$ contribution. Then, for the $e'_{\pm}$$^2$ state (Fig. 3(d)), the FM superexchange due to the above hoppings gains the energy against the AF by ($4t_{0}^{2}/U$)$\cdot$($2J_H/U$). But for the $a_{1g}$$^1$$e'_{-}$$^1$ ground state (Fig. 3(e)), the corresponding energy gain is more than doubled, ($10t_{0}^{2}/U$)$\cdot$($2J_H/U$). This could be a major reason why the $a_{1g}$$^1$$e'_{-}$$^1$ ground state has a much enhanced FM superexchange as compared to  $e'_{\pm}$$^{2}$, as shown in the above LSDA+SOC+U calculations.

Note that the $a_{1g}$$^1$$e'_{-}$$^1$ ground state has a local V$^{3+}$ spin moment of 1.88 $\mu_{\rm B}$ and an antiparallel orbital moment of --1.08 $\mu_{\rm B}$ along the $z$-axis, i.e., the SOC aligns the magnetic moment along the $z$-axis via the strong SIA, thus producing the perpendicular magnetic anisotropy and the resulting Ising magnetism.
Here we assume the spin Hamiltonian
\begin{equation}
H = -\frac{J}{2} \sum_{i,j}{\overrightarrow{S_{i}} \cdot \overrightarrow{S_{j}}} - D\sum_{i}{(S_{i}^{z})^{2}} - \frac{J'}{2}\sum_{i,j}S_{i}^z \cdot S_{j}^z,
\end{equation}
where the first term describes the Heisenberg isotropic exchange (FM when $J>$0), the second term is the SIA with the easy magnetization $z$ axis (when $D>$0), and the last term refers to the anisotropic exchange (the easy $z$ axis when $J'>$0). For VI$_3$ monolayer, the sum over $i$ runs over all V$^{3+}$ atoms with $S$=1 in the honeycomb lattice, and $j$ over the three first nearest V$^{3+}$ neighbors of each $i$. We have estimated the three magnetic parameters $J$, $D$ and $J'$, by calculating four different magnetic states, FM and AF with the perpendicular or in-plane magnetization, see Table I. The total energy results allow us to estimate $J$=6.07 meV, $D$=15.9 meV, and $J'$=0.67 meV. 
We see that the perpendicular magnetic anisotropy of the VI$_3$ monolayer arises predominantly
from the $D$-term, i.e., from the strong Ising-type SIA. This is one of the main results of this Letter. With these three parameters, our Monte Carlo simulations find that $T_{\rm C}$ of the VI$_3$ monolayer would be 40 K with the $J$-$J'$ contributions, 68 K with the $J$-$D$ contributions, and 74 K with the $J$-$D$-$J'$ contributions. Therefore, the $D$ contribution is about 5 times stronger than $J'$ in stabilizing the 2D FM order.

\begin{figure}
\includegraphics[width=7.5cm]{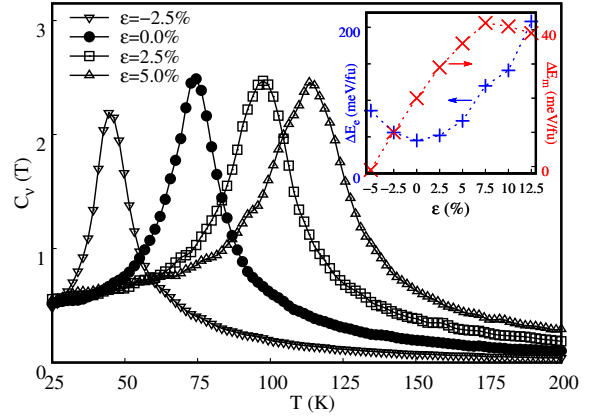}
 \caption{Monte Carlo simulations of the specific heat of the VI$_{3}$ monolayer under strains. The inset shows the relative stability of the $a_{1g}$$^1$$e'_{-}$$^1$ ground state against the $e'_{\pm}$$^{2}$ state (blue curve), and the FM stability of the $a_{1g}$$^1$$e'_{-}$$^1$ ground state against the AF state (red curve).
 }
\end{figure}

The above results remain largely unchanged when we carry out a structural optimization for the VI$_3$ monolayer, see Table I. The only significant change is that the insulating $a_{1g}$$^1$$e'_{-}$$^1$ ground state, at the theoretical equilibrium lattice constant $a$=$b$=6.70~\AA~(as compared with the experimental bulk value $a$=$b$=6.88~\AA~\cite{Kong_2019}), becomes even more stable than the $e'_{\pm}$$^2$ state, by 45.5 meV/fu. Furthermore, we study a biaxial strain effect on the FM of VI$_3$ monolayer. Our results show that the $a_{1g}$$^1$$e'_{-}$$^1$ ground state remains robust against the strains on the optimized lattice, see the blue curve in the inset of Fig. 4. The FM superexchange strength rises in the feasible tensile strain (e.g., up to 5\%) but decreases in the compressive strain, see the red curve. Then, assuming a feasible biaxial tensile strain, and using the increasing FM superexchange strength and Ising magnetism associated with the robust $a_{1g}$$^1$$e'_{-}$$^1$ ground state, we carry out Monte Carlo simulations to estimate $T_{\rm C}$ in the 2D Ising FM VI$_3$ monolayer. As seen in Fig. 4, $T_{\rm C}$ increases from 70 K for the bare VI$_3$ monolayer to 90 K at 2.5\% strain, and to 110 K at 5\% strain (but decreases to 45 K at --2.5\% strain). Therefore, VI$_3$ monolayer could be the first Ising type 2D FM with a pretty high $T_{\rm C}$, particularly under a biaxial tensile strain. This prediction calls for a prompt experimental verification.

The present case, VI$_3$, in comparison with the extensively studied CrI$_3$~\cite{Huang_2017,Huang_2018_nature,Jiang_2018,Sun_2019,Huang_2018_jacs,
Gibertini_2019,Song_2018,Mcguire_2015,Wang_2016,Lado_2017,Kim_2019},
demonstrates the importance of orbital degrees of freedom (orbital degeneracy)
in promoting novel exotic properties of 2D magnets. Whereas in CrI$_3$ without orbital degeneracy the SIA is negligible and the FM ordering in single layers
is due to the weak exchange anisotropy~\cite{Lado_2017,Kim_2019},
in the presence of orbital freedom as in VI$_3$ we get the significant Ising character
already at a single-ion level. Thus we can
propose that using orbital degrees of freedom may strongly enrich
magnetic properties of, in particular, 2D magnets.
Considering the common honeycomb structure of layered materials, the moderate (stronger) trigonal crystal field of the local octahedral $3d$ ($4d$) transition metal ions with partially filled $t_{2g}$ shells may favor a large orbital moment (in principle $L_z$=1 in size) and thus a strong perpendicular SIA, e.g., for high-spin $3d^2$ ($S$=1, $a_{1g}$$^1$$e'^1$), $3d^6$ ($S$=2, $e'^3$$a_{1g}$$^1$$e^2$), and $3d^7$ ($S$=3/2, $a_{1g}$$^2$$e'^3$$e^2$), and for $4d^2$ ($S$=1, $a_{1g}$$^1$$e'^1$) and low-spin $4d^4$ ($S$=1, $e'^3$$a_{1g}$$^1$). Here the $S$=1/2 ionic states are not listed as their magnetic coupling may not be sufficiently strong due to the small spins. The $5d$ ionic states are not discussed here as they are often nonmagnetic or weakly spin polarized due to the strong crystal field and significant covalency with the ligands. Then one could make use of these $S\ge1$ ionic states with strong SIA, as well as the above superexchange pictures associated with different orbitals, to search more 2D Ising magnets.


In summary, 2D FM materials are desirable for spintronics. Here we have demonstrated, through DFT calculations (LSDA+SOC+U), crystal field level analyses and superexchange pictures, that VI$_3$ monolayer has the $a_{1g}$$^1$$e'_{-}$$^1$ ground state with a significant SIA and enhanced FM couplings. This unique state produces an Ising type FM. Our results well account for the experimental hard perpendicular FM in bulk VI$_3$. Moreover, we predict, based on the MC simulations, that VI$_3$ monolayer could well be the first 2D Ising FM with a pretty high $T_{\rm C}$, particularly under a biaxial tensile strain. Furthermore, we suggest a group of spin-orbital states with a strong SIA which may help to search more 2D Ising magnets.


This work was supported by the NSF of China (Grants No. 11674064, No. 11704153, and No. 11474059) and by the National Key Research and Development Program of China (Grant No. 2016YFA0300700). D. I. K. was supported by the Deutsche Forschungsgemeinschaft through SFB 1238 (project number 277146847).

\nocite{xyz}

\bibliography{VI}

\begin{appendix}
\setcounter{figure}{0}
\setcounter{table}{0}
\renewcommand{\thefigure}{S\arabic{figure}}
\renewcommand{\thetable}{S\arabic{table}}
\section{Supplementary Material to ``VI$_{3}$: a 2D Ising ferromagnet''}

\section{I. Computational Details}

In our DFT calculations, a 20-\AA~ thick slab was used to model VI$_3$ monolayer. The experimental lattice parameters were used and the structural optimization was also carried out, and two sets of results turn out to be very similar as seen in the main text. The global coordinates are used, with the $z$-axis parallel to the crystallographic $c$-axis, i.e., along the [111] direction of the local VI$_6$ octahedra, see Fig. 1 in the main text. The muffin-tin sphere radii were chosen to be 2.2 Bohr for V and 2.5 Bohr for I. The plane wave expansion of the interstitial wave functionals was set to be 12 Ry. The Brillouin zone integration was performed over $12\times12\times1$ $k$-mesh.

\section{II. Cleavage Energy}

To explore the possibility to exfoliate a monolayer from the bulk VI$_{3}$, we have calculated the total energy of a VI$_3$ bilayer as a function of the interlayer distance (see Fig. S1), using DFT+vdW corrections within the Grimme's approach. The calculated cleavage energy of 0.27 J/m$^{2}$ is well comparable to those for other 2D materials such as CrI$_{3}$ (0.30 J/m$^{2}$), Cr$_{2}$Si$_{2}$Te$_{6}$ (0.35 J/m$^{2}$), and Cr$_{2}$Ge$_{2}$Te$_{6}$ (0.38 J/m$^{2}$). Therefore, an exfoliation of VI$_{3}$ monolayer is likely.

\begin{figure}[h]
\includegraphics[width=7cm]{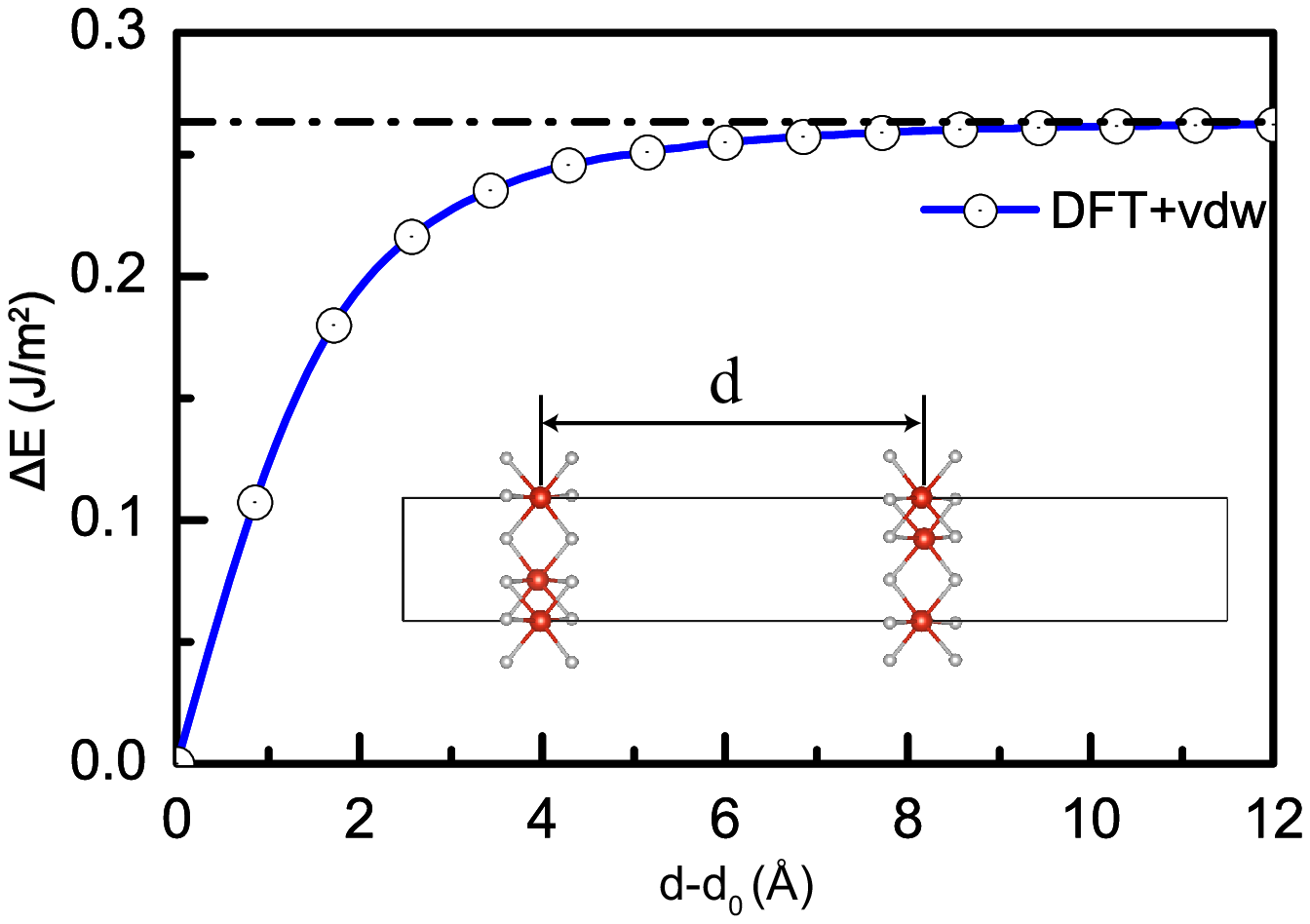}
 \caption{The relative total energy calculated as a function of the distance between two VI$_3$ monolayers with a reference to the experimental vdW distance $d_0$. The cleavage energy is estimated to be 0.27 J/m$^2$.
 }
\end{figure}

\section{III. LDA, LSDA, and LSDA+U results}

We carry out spin-restricted LDA calculations to estimate the crystal (ligand) field splitting in VI$_{3}$ monolayer. The calculated density of states (DOS) results are shown in Fig. S2. A  $t_{2g}$-$e_g$ like octahedral crystal field splitting of more than 1 eV makes the  empty $e_g$ states of no concern. Note that in the actual trigonal crystal field, the $t_{2g}$ triplet splits into the nearly degenerate $a_{1g}$ singlet and $e'_{\pm}$ doublet, both of which are on  average 1/3 filled to fulfill the formal V$^{3+}$ $3d^2$ state. When the spin-polarized LSDA calculations are performed, we obtain a FM metallic solution (Fig. S3) with the total spin moment of 2 $\mu$$_{\rm B}$/fu indicative of the formal V$^{3+}$ $S$=1 state. The V$^{3+}$ ion has the local spin moment of 1.85 $\mu_{\rm B}$ (see Table S1), and due to the strong covalency, each iodine becomes negatively spin polarized and has a local spin moment of --0.05 $\mu$$_{\rm B}$, and the interstitial region also contributes a large spin moment of 0.3 $\mu$$_{\rm B}$/fu. Obviously, this metallic solution contradicts the experimental FM insulating behavior. Note that the $t_{2g}$-like bandwidth is less than 1 eV (Figs. S2 and S3) and should be smaller than any realistic U value of about 3-4 eV for V $3d$ electrons. Therefore, the electronic correlations  should be taken into account, as done in the following +U calculations.

\begin{figure}[t]
\includegraphics[width=6cm]{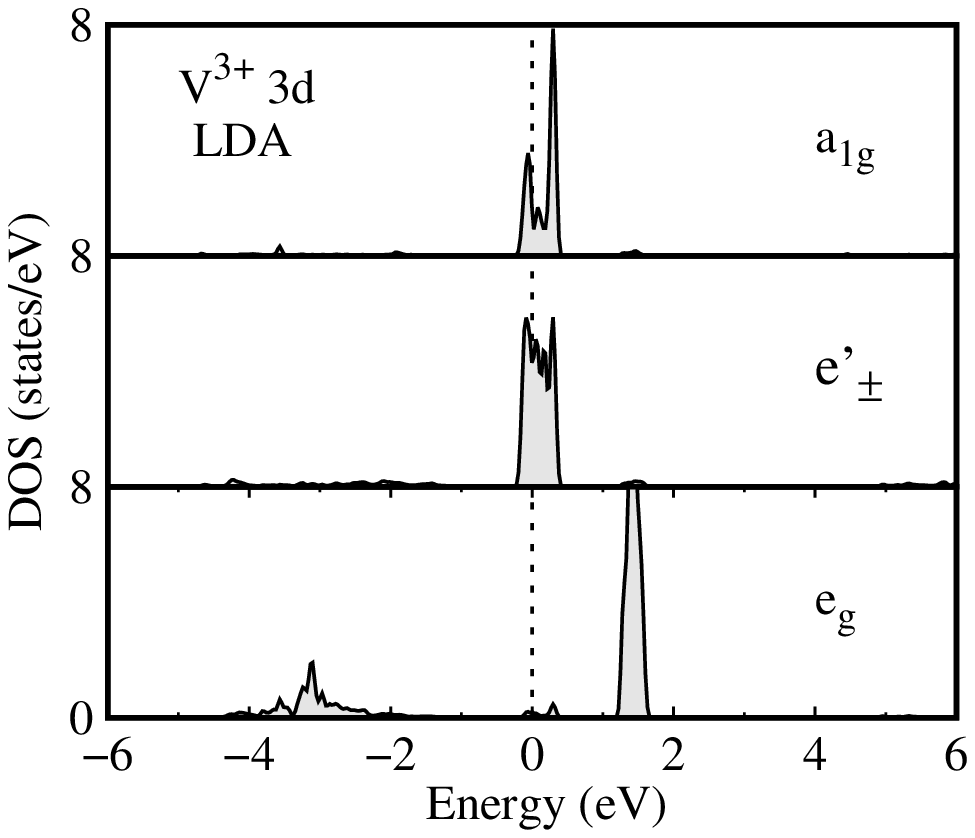}
 \caption{V $3d$ DOS for VI$_{3}$ monolayer by LDA. Fermi level is set at zero energy.
}
\end{figure}

\begin{figure}[!htp]
\includegraphics[width=6cm]{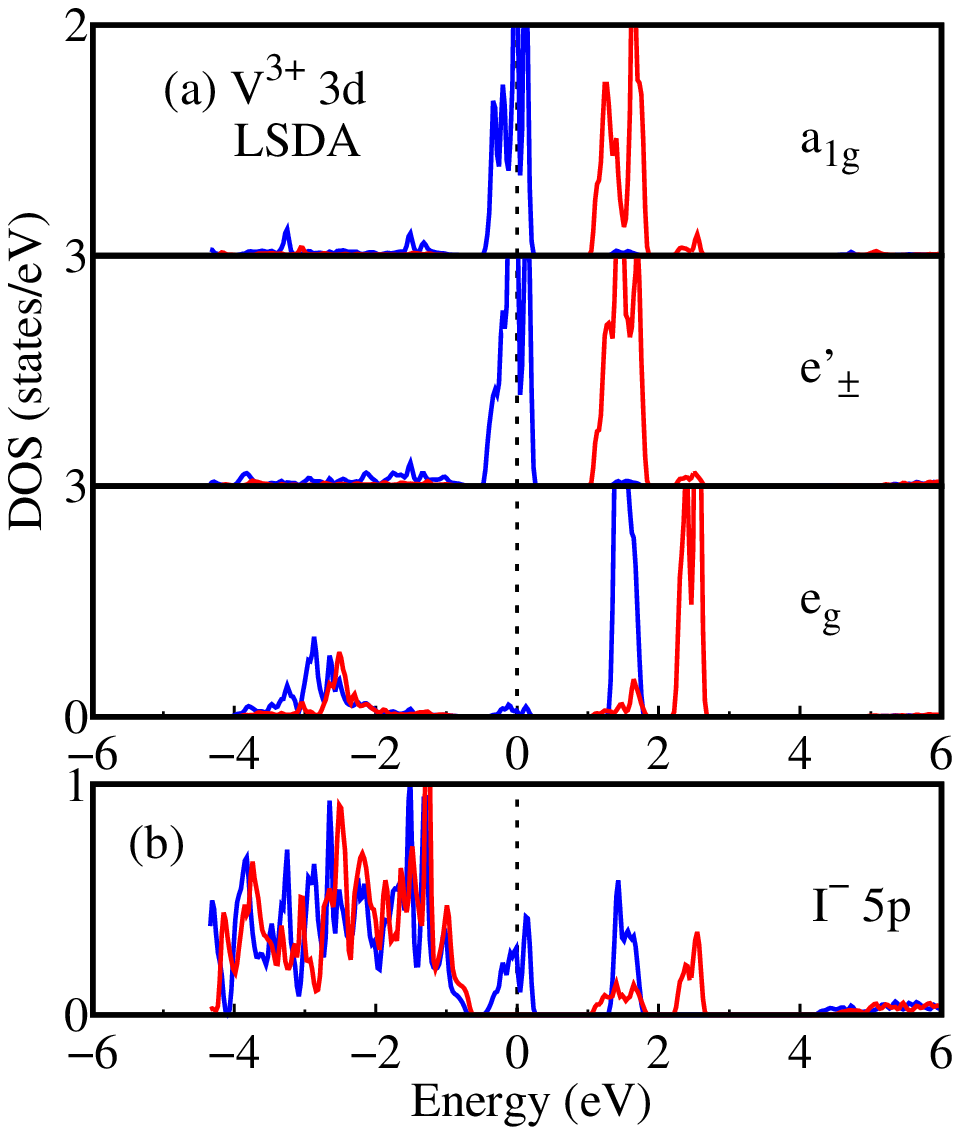}
 \caption{V $3d$ and I $5p$ DOS by LSDA. The blue (red) curves stand for the up (down) spin channel. Fermi level is set at zero energy.
}
\end{figure}

\renewcommand\arraystretch{1.3}
\begin{table}[!htp]
  \caption{Relative total energies $\Delta$E (meV/fu) and local spin moments ($\mu_{\rm B}$) for the V$^{3+}$ ion. The corresponding data for the fully relaxed structures are listed in the round brackets.
}
  \label{tb1}
\begin{tabular}{lccc}
\hline\hline
VI$_{3}$ monolayer  &   & $\Delta$E & M$_{\rm spin}$     \\ \hline
LSDA & FM      & 0.0 (0.0)    & 1.85 (1.77)   \\
LSDA+U ~$a_{1g}$$^{1}$$e'_{\pm}$$^{1}$~~  & FM      & 229.6 (193.7)   & 2.02 (1.98) \\
LSDA+U ~$e'_{\pm}$$^{2}$~~ & FM    & 0.0 (0.0) & 1.82 (1.77)               \\
& AF    & 7.2 (7.2) & 1.80 (1.75)           \\
\hline\hline
 \end{tabular}
\end{table}
\begin{figure}[!t]
\includegraphics[width=6cm]{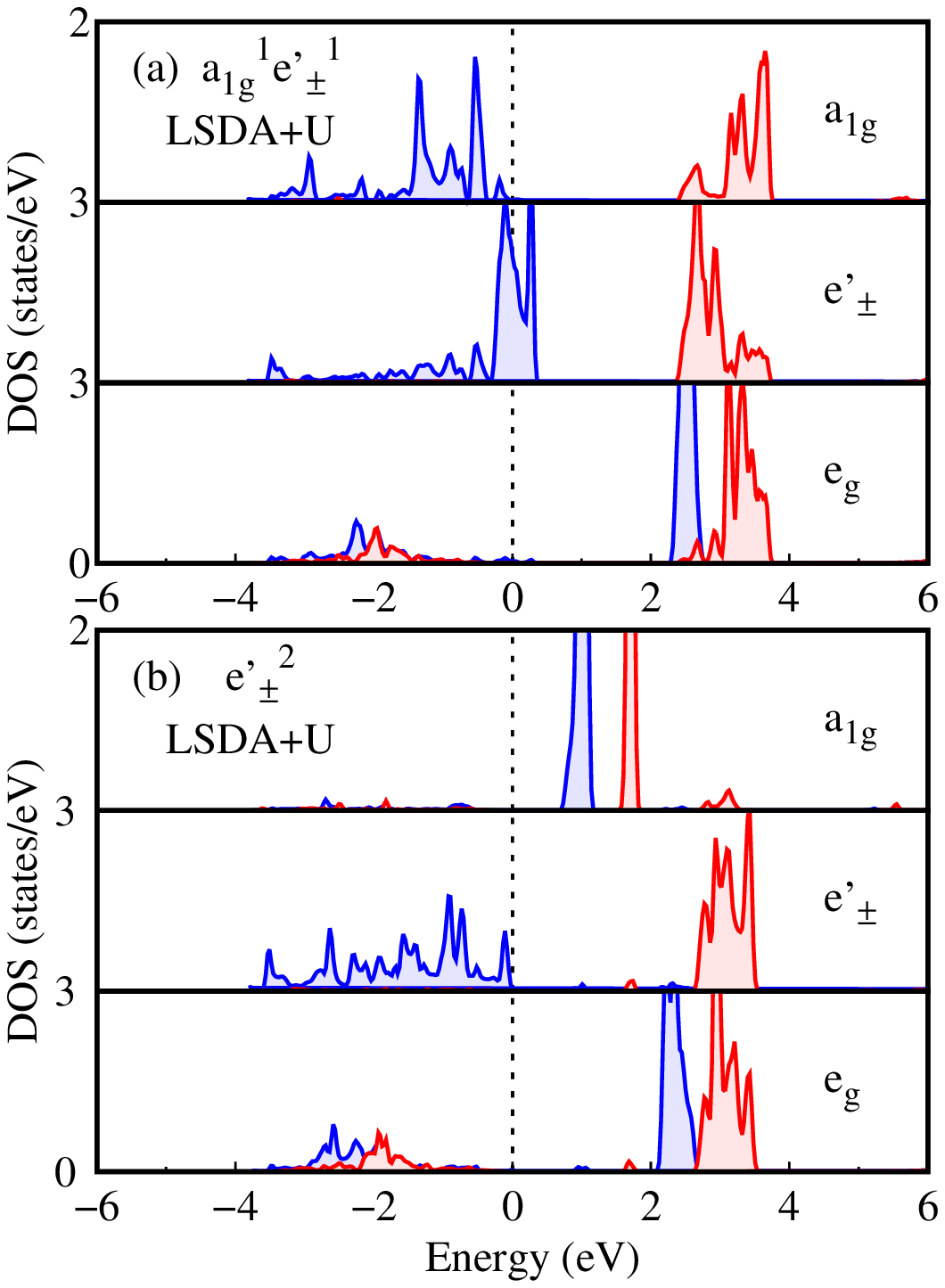}
 \caption{(a) The metallic $a_{1g}$$^1$$e'_{\pm}$$^1$ state and (b) the insulating $e'_{\pm}$$^2$ state by LSDA+U. The blue (red) curves stand for the up (down) spin channel. Fermi level is set at zero energy.
}
\end{figure}

Our LSDA+U calculations give two different solutions: the metallic one ($a_{1g}$$^1$$e'_{\pm}$$^1$) with the half-filled $e'_{\pm}$$^1$ state and the insulating one with the fully occupied $e'_{\pm}$$^2$ configuration, see Fig. S4. The former solution has a sharp DOS peak at the Fermi level, and therefore it meets the scenario of a Stoner FM instability. As a result, this solution may be named an itinerant FM, which is, however, in disagreement with the experimental FM insulating behavior. Actually, our LSDA+U calculations find that this solution with the high DOS peak at the Fermi level is much less stable than the insulating $e'_{\pm}$$^2$ state by 229.6 meV/fu, see Table S1. Moreover, the insulating $e'_{\pm}$$^2$ state turns out to be more stable in FM state than in the AF state by 7.2 meV/fu.

\end{appendix}
\end{document}